\documentclass[twocolumn,amsmath,prd,aps,aas_macros,tightenlines,superscriptaddress,preprintnumbers, nofootinbib]{revtex4}

\usepackage{graphicx, color}
\usepackage{dcolumn}
\usepackage{epsfig}
\usepackage{amssymb}
\usepackage{amsmath}

\def\fun#1#2{\lower3.6pt\vbox{\baselineskip0pt\lineskip.9pt
  \ialign{$\mathsurround=0pt#1\hfil##\hfil$\crcr#2\crcr\sim\crcr}}}

\usepackage{graphicx}\usepackage{dcolumn}
\usepackage{bm}
\usepackage{amssymb}
\usepackage{amsmath}
\usepackage{hyperref}
\usepackage{natbib}

                              \newlength{\strikewidth}
                              \newlength{\strikelength}
\newcommand{\vect}[1]{\mbox{\boldmath${#1}$}}
\newcommand{\nhat}{\hat{{\vect{n}}}}
\newcommand{\mhat}{\hat{{\vect{m}}}}
\newcommand{\khat}{\hat{{\vect{k}}}}

\def\mnras{MNRAS}
\def\apjl{ApJ}   

\begin{document}

\title{CMB Isotropy Anomalies and the Local Kinetic Sunyaev-Zel'dovich Effect}
\author{Hiranya V. Peiris}\email{h.peiris@ucl.ac.uk}
\affiliation{Institute of Astronomy and Kavli Institute for Cosmology, University of Cambridge, Cambridge CB3 0HA, U.K.}
\affiliation{Department of Physics and Astronomy, University College London, London WC1E 6BT, U.K.}
\author{Tristan L. Smith}\email{tlsmith@berkeley.edu}
\affiliation{Berkeley Center for Cosmological Physics, University of California, Berkeley, CA 94720 U.S.A.}

\begin{abstract}
Several anomalies have been identified which may imply a breakdown of the statistical isotropy of the cosmic microwave background (CMB). In particular, an anomalous alignment of the quadrupole and octopole and a hemispherical power asymmetry have increased in significance as the data have improved. There have been several attempts to explain these observations which explore isotropy breaking mechanisms within the early universe, but little attention has been given to the possibility that these anomalies have their origin within the local universe.  We explore such a mechanism by considering the kinetic Sunyaev-Zel'dovich effect due to a gaseous halo associated with the Milky Way. Considering several physical models of an anisotropic free electron optical depth contributed by such a halo, we find that the associated screening maps of the primordial anisotropies have the necessary orientations to affect the anomaly statistics very significantly, but only if the column density of free electrons in the halo is at least an order of magnitude higher than indicated by current observations.
\end{abstract}
\maketitle

\section{Introduction}

High-precision measurements of the cosmic microwave background (CMB) anisotropies and the agreement of these observations with our theoretical expectations has elevated cosmology to a precision science.  The observed anisotropies have been measured to follow isotropic, Gaussian statistics with an almost scale-invariant power-spectrum.  However, as observations have improved, anomalies which might indicate deviations from the theoretical expectations have increased in significance.  Prosaic explanations for the observed anomalies may yet be found as our understanding of the systematics of the observations improves, but it is also of interest to explore whether these anomalies point us towards exotic and unexpected physics.  

One such anomaly is the apparent breakdown of statistical isotropy that has been reported in the CMB fluctuations at the largest observable scales \cite{Copi:2003kt,deOliveiraCosta:2003pu,Eriksen:2003db,Schwarz:2004gk,Land:2005ad} measured by the Wilkinson Microwave Anisotropy Probe (WMAP).  Two major observations\footnote{A quadrupolar anisotropy of the inferred primordial power spectrum has also been reported \cite{Groeneboom:2008fz, Hanson:2009gu,Groeneboom:2009cb}, which is aligned with the ecliptic. However, this signal shows significant variations between different WMAP differencing assemblies at the same frequency \cite{Hanson:2009gu}, pointing to an experimental systematic which has not been taken into account in these analyses.} that are suggestive of this statistical anisotropy refer to particular special directions in the sky.  The first is the planarity of the quadrupole and octopole and their mutual alignment; this plane is also roughly orthogonal to the CMB dipole direction  \cite{Land:2005ad,Copi:2003kt,deOliveiraCosta:2003pu}.  The second is an asymmetry in the amplitude of the power spectrum between the northern and southern ecliptic hemispheres \cite{Eriksen:2003db} which has also been observed in Cosmic Background Explorer (COBE) data but at a lower significance  \cite{Eriksen:2003db}. 

With the evidence for these anomalies increasing as the data improves \cite{Land:2006bn,Hansen:2008ym} many studies have proposed modifications of early-universe physics in order to generate a violation of statistical isotropy (see Refs.~\cite{Ackerman:2007nb,Erickcek:2008sm,Erickcek:2008jp,Erickcek:2009at} for a small selection).  There have also been discussions of how a violation of statistical isotropy would affect other observations assuming that its source is primordial \cite{Dvorkin:2007jp,Hirata:2009ar,Frommert:2009qw, Groeneboom:2009cb}, enabling consistency tests of those hypotheses.  

However, soon after these anomalies were discovered, it was noticed that the directions associated with them corresponded to special directions within the local universe: four of the planes associated with the quadrupole and octopole are orthogonal to the ecliptic, the remaining octopole plane is orthogonal to the Galactic plane, and the hemispherical asymmetry is aligned with the ecliptic.  The heuristic connection between these CMB anomalies and our local environment has generated only a handful of quantitative attempts to connect the two, including possible foregrounds associated with the heliosphere \cite{Frisch:2007kz}, the kinetic Sunyaev-Zel'dovich effect of free electrons in the Galactic disk \cite{Hajian:2007xi,Waelkens:2007wn}, the thermal Sunyaev-Zel'dovich effect of the local universe \cite{Suto1996, Abramo:2006hs}, the Rees-Sciama effect of the local superclusters \cite{Maturi:2007xr}, and the Integrated Sachs-Wolfe effect of the low-redshift universe \cite{Francis:2009pt}. This is an area which deserves greater attention, since a very local signal can affect the largest observable scales.

In this paper, we explore another possible local origin of the anomalies by investigating how the scattering of CMB photons by free electrons diffusely distributed within the Milky Way halo through the kinetic Sunyaev-Zel'dovich (kSZ) effect may introduce {\it an anisotropic contamination} of the primordial signal and help explain the origin of the anomalies and their alignment with special directions in the local universe.  The effects of the kSZ from the Galactic halo on the CMB has been recently discussed in Ref.~\cite{Birnboim:2009ne}.  However Ref.~\cite{Birnboim:2009ne} did not explore how the kSZ may impact these anomalies and instead concentrated on whether the local kSZ may produce a foreground that must be subtracted in order to extract the primordial signal.  

The prediction of an extended gaseous halo within the Milky Way is quite robust.  Models of galaxy formation for halos with masses $M \gtrsim 10^{12} M_{\odot}$ predict infalling gas is shock-heated to the virial temperature (around $10^{6}-10^{7}$ K for the Milky Way) and remains in hydrostatic equilibrium until it is able to cool and condense to form stars \cite{Keres:2004cq,Birnboim:2003xa}.  For a Milky Way-sized halo it is expected that the baryonic mass fraction follows the cosmic average, $f \sim 0.1$.  By subtracting the total baryonic mass in stars and gas in the Galaxy, a reasonable estimate of the baryonic mass in an extended, hot, gaseous halo is $\sim 5 \times 10^{10}\ M_{\odot}$\cite{Birnboim:2009ne}.  Observations of OVII and OVIII X-ray absorption have also indicated the existence of an extended hot gaseous halo \cite{2007ApJ...669..990B} associated with the Milky Way.  Furthermore, observations of pulsars yield estimates for the column density of free electrons and lead to a free-electron fraction within the Milky Way halo which is consistent with a hot gaseous halo with a column density of $\sim 10^{21}\ {\rm cm}^{-2}$ \cite{Taylor:1993my}.  Given that the Milky Way is moving relative to the CMB with a velocity $v/c \sim 10^{-2}$, the local kSZ may produce a signal around 1 $\mu$K.  

In \S~\ref{sec:2}  we point out that any anisotropic optical depth to Thomson scattering off local electrons, coupled through the kSZ effect with the dipole of the local electrons with respect to the CMB, induces an anisotropic imprint with a black-body spectrum. Its alignment is determined by the dipole direction and the anisotropy of the distribution. Given that the kSZ is expected to be subdominant to the cosmological signal, in \S~\ref{sec:impact} we point out the disproportionate impact on anomaly statistics of small signals. We make no judgement on the usefulness or otherwise of such {\sl a posteriori} anomaly statistics (for a post-WMAP 7 year discussion of this point, see Ref.~\cite{Bennett:2010jb}). Instead, we take at face value anomaly statistics which have been invoked in a large body of literature, and study the impact of the kSZ signal on their significance. In \S~\ref{sec:doppler_dip} we derive the coupling of an anisotropic electron screen to the Doppler dipole, and consider several physical models of the form of the anisotropic electron distribution in \S~\ref{sec:models}. We compute the impact of these halo kSZ models on anomaly statistics in \S~\ref{sec:stats}, and discuss the results in \S~\ref{sec:discuss}.

\section{Effect of local kSZ on CMB anisotropies}\label{sec:2}

A simple way to understand how an anisotropic optical depth leads to additional anisotropy in the CMB is to note that along a given line of sight, $\hat{\vect{n}}$, with optical depth $\tau(\nhat)$, the fraction of photons which scatter out of the line of sight is given by
\begin{equation}
\left[1-e^{-\tau({\hat{\bf n}})}\right]\left[\bar{T}+ \Delta T(\nhat)\right],
\end{equation}
where $\Delta T(\nhat)$ is the temperature anisotropy along that line of sight.  In addition, there are photons that scatter in to the line of sight isotropically from every other part of the sky; thus, they contribute the mean temperature, $\bar{T}$, and the observed temperature anisotropy is \cite{Haiman:1999me}
\begin{eqnarray}
T_{\rm obs} &=& (\bar{T} + \Delta T) - \left[1-e^{-\tau(\hat{\bf n})}\right]\left[\bar{T} + \Delta T\right] \nonumber \\
& & + \bar{T} \left[1-e^{-\tau(\hat{\bf n})}\right] \nonumber \\
&=& \bar{T} + \Delta T(\nhat) e^{-\tau(\hat{\bf n})}.
\end{eqnarray}
From this simple calculation we can see that an anisotropic optical depth will couple with any inherent temperature anisotropies and thereby modulate the observed anisotropy.  

A more rigorous way to derive the same result is to consider the Boltzmann equation which dictates the evolution of the photon distribution function.  In cosmology the application of the Boltzmann equation leads to the usual evolution equations for the photon distribution function within a perturbed Friedmann-Roberson-Walker (FRW) universe \cite{Dodelson:2003ft,Challinor:2009tp}.  It can also be used to describe the evolution of the photon distribution function as the photons pass through the Milky Way's halo.  Denoting perturbations to the photon temperature by $\bar{T} \Theta(t,\vect{x},\nhat)$ the Boltzmann equation gives 
\begin{eqnarray}
\frac{d \Theta}{dt} = \frac{\partial \Theta}{\partial t} - \nhat \cdot \vect{\nabla} \Theta  \hspace{1.8in} \label{eq:Boltzmann1} \\
=  \sigma_T n_e\bigg(- \Theta +  \frac{3 }{16 \pi} \int d^2 \mhat \ \Theta(\mhat) \left[ 1+ (\nhat \cdot \mhat)^2\right] \nonumber \\
  - \nhat \cdot \vect{v}_b/c\bigg), \nonumber 
\end{eqnarray}
where the first term describes scattering out of the line of sight, the second term is due to scattering into the line of sight, the last term is the Doppler shift due to the bulk motion of the electrons (with velocity $\vect{v}_b$ in the CMB rest-frame). The Thomson scattering cross-section is $\sigma_T = 6.65 \times 10^{-25}\ {\rm cm^2}$. 

Any optical depth contributed by free electrons within the Milky Way halo will be much smaller than unity so we may solve Eq.~(\ref{eq:Boltzmann1}) perturbatively.  The zeroth order solution follows $d\Theta^{(0)}/dt=0$.  We then find the first order correction to this 
\begin{equation}
\frac{\partial \tilde{\Theta}^{(1)}(\vect{k})}{\partial t} - i k \mu \tilde{\Theta}^{(1)} (\vect{k}) = \int d^3 x\ e^{i {\bf k} \cdot {\bf x}}  \sigma_T n_e(\vect{x}) S^{(0)}(\vect{x}, \nhat),
\label{eq:Boltzmann2}
\end{equation}
where $\mu = \nhat \cdot \khat$ and $S^{(0)}$ is the term in parentheses in Eq.~(\ref{eq:Boltzmann1}) evaluated for the zeroth-order solution.  In the case of the CMB this term is approximately given by $S^{(0)} = -\nhat \cdot \vect{v}_b$ since $v_b/c = 6 \times 10^{-3}$ \cite{Kogut:1993ag} and $\tilde{\Theta}^{(0)} \sim 10^{-5}$. As can be verified by substitution, the solution to Eq.~(\ref{eq:Boltzmann2}) is then given by
\begin{equation}
\Theta^{(1)}(\nhat) = -\nhat\cdot \vect{v}_b\ \sigma_T \int_0^{\infty} ds\ n_e(s \nhat),
\end{equation}
where the integral extends along the line of sight.  The observed temperature pattern is then given by
\begin{eqnarray}
\Theta^{\rm obs} &=& \Theta^{\rm cos} - \nhat \cdot \vect{v}_b\ \sigma_T \int_0^{\infty} ds\ n_e(s \nhat) \nonumber \\
&=& \Theta^{\rm cos} - \left[\nhat \cdot \vect{v}_b\right] \tau(\nhat), 
\label{eq:obs_T}
\end{eqnarray}
where we have explicitly written $\Theta^{(0)} = \Theta^{\rm cos}$ and $n_e$ is the number density of electrons within the Milky Way halo. 
Defining $\mathcal{C}(\nhat) \equiv \sigma_T \int_0^{\infty} ds \ n_e(s \nhat)$, the amplitude of the second term in Eq.~(\ref{eq:obs_T}) can be written
\begin{eqnarray}
\nhat\cdot \vect{v}_b\ \sigma_T \int_0^{\infty} ds\ n_e(s \nhat) \hspace{1.5in} \nonumber \\
= 1.33\times 10^{-6} \frac{\bar{\mathcal{C}}}{10^{21}\ {\rm cm^{-2}}}\frac{\mathcal{C}(\nhat)}{\bar{\mathcal{C}}}\frac{\nhat \cdot \vect{v}_b}{600\ {\rm km/s}},
\label{eq:angle_avg}
\end{eqnarray}
where we have defined the angle averaged $\bar{\mathcal{C}} \equiv \int d^2 \hat{n}\ \mathcal{C}(\hat{n})/(4\pi)$.

Without any reference to a specific model for the optical depth, it is clear that the kSZ is capable of producing any modulation of the CMB that we may observe.  Letting $\tau(\nhat)$ denote the optical depth as a function of position on the sky, assuming some underlying primordial temperature anisotropies then the optical depth is determined by
\begin{equation}
\tau(\nhat) = \frac{\Theta^{\rm cos}-\Theta^{\rm obs}}{\nhat \cdot \vect{v}_b}.
\end{equation}

\section{Disproportionate impact of small signals on anomaly statistics} \label{sec:impact}

While, in principle, this effect can generate the observed large-angle isotropy anomalies, in practice (given the expected column density of free electrons and the dipole velocity of the CMB), estimates of the kSZ are at the $1$ $\mu$K level (compared to the $\pm 20$ $\mu$K of the quadrupole). Given that, at most, the kSZ can produce $\sim$ 10\% modulations of the primary CMB, it would appear as though its effect on the observed anomalies would be negligible.  However, as we now discuss, even a 10\% modulation can have a large effect on the inferred statistical significance of these anomalies.  Consider the ``angular momentum dispersion'' statistic \cite{deOliveiraCosta:2003pu, Gordon:2005ai, Dvorkin:2007jp} for quantifying the planarity of multipole $\ell$:
\begin{equation}
L^2_{\ell}(\nhat) = \frac{\sum_{m= -\ell}^{\ell} m^2 |a^{\rm obs}_{\ell m}|^2}{\ell^2 \sum_{m=-\ell}^{\ell}  |a^{\rm obs}_{\ell m}|^2}\ ,
\end{equation}
which, given the observed realization on the sky, is maximized in some direction $\nhat'$. The transformation between the general frame and the maximizing frame is given by 
\begin{equation}
a_{\ell m'} = \sum_{m=-\ell}^{\ell} D^\ell_{m m'} (\tilde\phi, -\theta, -\phi)\ a_{\ell m}\ , 
\end{equation}
where $D$ is the Wigner matrix corresponding to the appropriate rotation between the frames, and $\tilde\phi$ can take any value. The statistics given by the maximum values of $L^2_2$, $L^2_3$ and $L^2_{23}=L^2_2+L^2_3$ have been used in the literature to quantify the planarity of the quadrupole and octopole, and also to capture their mutual alignment, respectively. 

Table~\ref{tab:dispersionstats} shows the values and $p$-values for these statistics for the WMAP 7 year Internal Linear Combination (ILC7) foreground-cleaned map \cite{Jarosik:2010iu}, with and without subtracting the kinematic quadrupole (KQ). The $p$-values are computed from 10,000 isotropic realizations. The kinematic quadrupole induced by the dipole anisotropy  due to our motion relative to the CMB is at the level of $-1$ $\mu$K $\rightarrow$ $+3$ $\mu$K, whereas the CMB quadrupole is at the $-22$ $\mu$K $\rightarrow$ $+16$ $\mu$K level. The Table shows that the subtraction of the very subdominant KQ signal leads to a factor of $\sim 2.7$ decrease in the $p$-value of the most anomalous statistic, $L^2_{23}$. This effect is even more dramatic for the alternative WMAP 5 year Tegmark-Oliveira-Costa-Hamilton (TOH5) foreground-cleaned map \cite{deOliveiraCosta:2003pu}, where the maximum value of $L^2_{23}$ goes from $0.93$ to $0.96$ upon subtracting the KQ. The $p$-value corresponding to the latter is $0.0029$, making the statistic a factor of $5.5$ more unlikely under the null hypothesis of isotropy. This dramatic impact of subtracting the small KQ signal was also noticed in the WMAP first year data by Ref.~\cite{Copietal2006}, and is also present in the WMAP 5 year ILC (ILC5) map \cite{Hinshaw:2008kr}. For purposes of comparison, the $p$-value for the ILC5-KQ  $L^2_{23}=0.95$ is $0.0062$.

\begin{table}[!ht]
\caption{Values and $p$-values (computed from 10,000 isotropic realizations) for angular momentum dispersion statistics $L^2_2$, $L^2_3$ and $L^2_{23}$ (see text) for the WMAP 7 year Internal Linear Combination (ILC7) map, with or without subtracting the kinematic quadrupole (KQ).}
\begin{center}
\begin{tabular}{lcccc}
\hline
Statistic & ILC7  & ILC7  & ILC7-KQ & ILC7-KQ \\
	& maximum & $p$-value & maximum &  $p$-value \\
\hline
$L^2_2$ & $0.943$	& 0.66	&  $0.983$	&  0.38\\
$L^2_3$ & $0.919$ 	& 0.17	&  $0.926$	& 0.15 \\
$L^2_{23}$ & $0.931$ & 0.015 & $0.953$ & 0.0055 \\
\hline\\
\end{tabular}
\end{center}
\label{tab:dispersionstats}
\end{table}

\begin{figure}[!ht]
\includegraphics[scale=0.5]{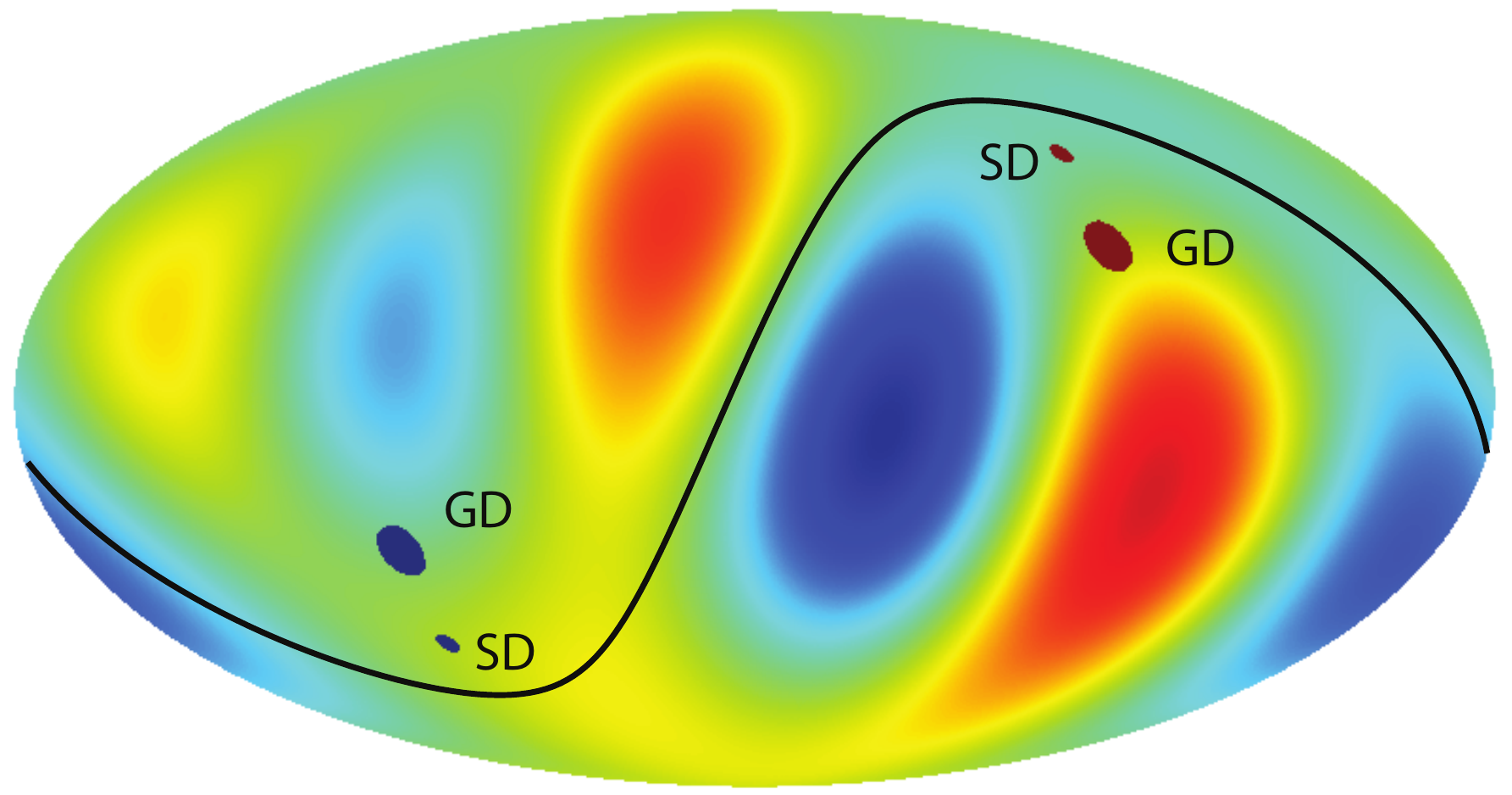}
\caption{The WMAP 5 year quadrupole and octopole, with the ecliptic plane (solid line), the Solar system dipole (SD), and the Galactic barycenter dipole (GD).}
\label{fig:directions}
\end{figure}

Given these observations, it is very important to quantify the impact of the kSZ effect of the Galactic halo for two reasons. First, the velocity of the Galactic barycenter with respect to the CMB points in the direction $(l,b) = (266.5^\circ, 29.1^\circ)$ \cite{COBEDipole1993}, which is quite close to the Solar CMB dipole direction, $(263.8^\circ, 48.2^\circ)$ and thus relevant for the orientations picked out by the large-angle isotropy anomalies (see Fig.~\ref{fig:directions}). Secondly, measurements of the column density of free electrons in the Galactic halo are not very precise and come primarily from theoretical arguments; observational limits are very difficult to obtain and the uncertainties on constraints are large \cite{Bregman:2009qh}. Thus, in the following calculations, we will consider a number of plausible physical models of the geometry of the halo free electron optical depth, but allow the magnitude of the effect to float somewhat, roughly at the level of the KQ amplitude, while keeping it greatly subdominant to the CMB quadrupole/octopole amplitudes.

 \section{Coupling to Doppler dipole \label{sec:doppler_dip}}
  
After processing through an anisotropic Thomson scattering screen, an anisotropic temperature field $\Theta$ gets screened as $\Theta(\nhat) \rightarrow \Theta(\nhat)\exp(-\tau(\nhat))$, as we have seen. Consider that the electron distribution consists of an isotropic part $\bar\tau$ and a (small) anisotropy $\tau^{\rm scr}(\nhat)$. This can be expanded as $e^{-\tau(\nhat)} \simeq e^{-\bar\tau}(1-\tau^{\rm scr}(\nhat))$, leading to an observed temperature anisotropy $\Theta^{\rm obs}(\nhat) = e^{-\bar\tau}\left[1-\tau^{\rm scr}(\nhat)\right] \Theta(\nhat)$, with the screened component of the temperature field $T^{\rm scr}(\nhat) = e^{-\bar\tau}\tau^{\rm scr}(\nhat)\Theta(\nhat)$.

Let us expand the anisotropic part of the local optical depth in spherical harmonics, 
\begin{equation}
\tau^{\rm scr}(\nhat) = \sum_{\ell_1\geq 1} \sum_{m_1} \tau_{\ell_1 m_1} Y_{\ell_1 m_1}  (\nhat),
\end{equation}
 which we will couple to the Doppler dipole of the CMB, 
\begin{equation}
T^{\rm dip}(\nhat) = A_{\rm d} Y_{1 0}  (\nhat).
\end{equation}
 The screened temperature field after passing through the anisotropic $\tau^{\rm scr}$, assuming this is much smaller than the isotropic part of the screening optical depth, is then given by
 \begin{equation}
T^{\rm scr}(\nhat) \simeq e^{-\bar\tau} \tau^{\rm scr}(\nhat) T^{\rm dip} (\nhat)
\end{equation}
Expanding $T^{\rm scr}$ in spherical hamonics, we obtain
\begin{eqnarray}
T^{\rm scr}_{\ell m} &=&  e^{-\bar\tau} \int d^2 \nhat\ \tau^{\rm scr}(\nhat) T^{\rm dip}(\nhat) Y^*_{\ell m} (\nhat), \nonumber \\
 &=& e^{-\bar\tau} (-1)^m \int d^2 \nhat\  \tau^{\rm scr}(\nhat) T^{\rm dip}(\nhat)  Y^*_{\ell m} (\nhat),  \nonumber \\
 &=& e^{-\bar\tau} (-1)^m  A_{\rm d}  \sum_{\ell_1 m_1} \tau_{\ell_1 m_1} \times \nonumber \\
 & &  \int d^2  \nhat\ Y_{\ell_1 m_1}  (\nhat) Y_{1 0}  (\nhat) Y^*_{\ell m} (\nhat) .
\end{eqnarray}
The right hand side of this equation can be expressed in terms of Wigner 3$j$ symbols using the Gaunt integral
\begin{eqnarray}
\int d^2 \nhat\  Y_{\ell_1 m_1} (\nhat) Y_{\ell_2 m_2}  (\nhat) Y_{\ell_3 m_3}  (\nhat) &=& \nonumber \\
 \sqrt{\frac{(2\ell_1+1)(2\ell_2+1)(2\ell_3+1)}{4\pi}} \times \nonumber \\
 \left( \begin{array}{ccc}
\ell_1 & \ell_2 & \ell_3 \\
0 & 0 & 0 \end{array} \right)
\left(\begin{array}{ccc}
\ell_1 & \ell_2 & \ell_3 \\
m_1 & m_2 & m_3 \end{array} \right),
\end{eqnarray}
leading to 
\begin{eqnarray}
\label{eq:couple}
T^{\rm scr}_{\ell m} &=& e^{-\bar\tau} (-1)^m  A_{\rm d} \sum_{\ell_1 m_1} \tau_{\ell_1 m_1}  
 \times \nonumber  \\
& & \sqrt{\frac{3(2\ell_1+1)(2\ell+1)}{4\pi}} \left( \begin{array}{ccc}
\ell_1 & 1 & \ell \\
0 & 0 & 0 \end{array} \right)
\left(\begin{array}{ccc}
\ell_1 & 1 & \ell \\
m_1 & 0 & -m \end{array} \right), \nonumber \\
&=&  e^{-\bar\tau} (-1)^m  A_{\rm d} \sum_{\ell_1} \tau_{\ell_1 m}   
\sqrt{\frac{3(2\ell_1+1)(2\ell+1)}{4\pi}} \times \nonumber  \\
& & \left( \begin{array}{ccc}
\ell_1 & 1 & \ell \\
0 & 0 & 0 \end{array} \right)
\left(\begin{array}{ccc}
\ell_1 & 1 & \ell \\
m & 0 & -m \end{array} \right),
\end{eqnarray}
where the second expression has made use of the symmetry property of the Gaunt integral that $m_1 + m_2 + m_3 = 0$. The first 3$j$ symbol in Eq.~(\ref{eq:couple}) enforces the triangle condition $\ell_1=\ell \pm 1$, so finally we obtain
\begin{equation}
\label{eq:tscreen}
T^{\rm scr}_{\ell m} = e^{-\bar\tau} A_{\rm d} \sqrt{\frac{3}{4\pi}} \left[ C_{\ell-1, m} \tau_{\ell-1, m}  + C_{\ell+1, m} \tau_{\ell+1, m} \right], 
\end{equation}
where
\begin{eqnarray}
C_{\ell-1, m} &=&   \sqrt{\frac{\ell^2 - m^2}{4\ell^2-1}}\\
C_{\ell+1, m} &=&  \sqrt{\frac{(1+\ell)^2 - m^2}{(1+2\ell)(3+2\ell)}} .
\end{eqnarray}
 Remember that in these expressions, the $\tau_{\ell'' m''}$ starts at $\ell''=1$. So, $T_{0m} \propto C_{1m}\tau_{1m}$, $T_{1m} \propto C_{2m}\tau_{2m}$, $T_{2m} \propto (C_{1m}\tau_{1m} + C_{3m}\tau_{3m})$, and so on. 

\section{Models for anisotropic free electron distributions} \label{sec:models}

In this section we consider how an extended gaseous partially ionized halo couples the CMB to the Doppler dipole through the kSZ effect, first gaining physical intuition using an analytic toy model of an anisotropic optical depth distribution, and then considering physical models.
 
\subsection{Simple geometries for the electron distribution}
 
Let us develop a feel for some simple, plausible electron distributions. Consider $\tau(\theta,\phi)$, a smooth function defined on the unit sphere, with $(0 \leq \theta \leq \pi, 0 \leq \phi \leq 2\pi)$, which we want to expand in terms of spherical harmonics as 
 \begin{equation}
 \tau(\theta,\phi) = \sum_{\ell=0}^\infty \sum_{m=-\ell}^\ell a_{\ell m} Y_{\ell m}(\theta, \phi) . 
 \end{equation}
 The spherical harmonic coefficients for $\tau$ are given as usual by
 \begin{equation}
 a_{\ell m} = \int_0^{2\pi} \int_0^\pi d\phi\ d\theta \sin \theta\ \tau(\theta, \phi) Y^*_{\ell m} (\theta, \phi). 
 \end{equation}
 A sphere only has the non-zero coefficient $a_{00}$ so is not useful for our purpose. Consider the ellipsoid centered on the origin and aligned with the coordinate axes, 
 \begin{equation}
 \left(\frac{x}{a}\right)^2 + \left(\frac{y}{a}\right)^2 + \left(\frac{z}{c}\right)^2 = 1, 
 \end{equation}
 where as usual, $x= \tau \sin \theta \cos \phi$, $y=\tau \sin\theta \sin\phi$, $z=\tau \cos \theta$. Thus for this distribution,
 \begin{equation}
 \tau(\theta,\phi) = ac\left( a^2 \cos^2 \theta + c^2 \sin^2 \theta \right)^{-1/2}.
 \end{equation}
 This function only depends on $\theta$, so the only non-zero spherical harmonic coefficients have $m=0$. Further, since $\tau(\theta,\phi)$ is even in $\theta$, only $a_{\ell 0}$ where $\ell$ is even are non-zero. This is interesting because it says that if the anisotropy in $\tau$ is axisymmetric in the dipole frame, $T^{\rm scr} = T^{\rm scr}_{\ell 0}$ where only $\ell=$ {\em odd} multipole couplings would survive. While in the real world, any local free electron distribution is unlikely to be anisotropic in a axisymmetric way in the CMB dipole frame, we can see that in general there will be an asymmetry between odd and even multipoles in the power spectrum of the screening field. This analytic argument thus explains the ``sawtooth" pattern of the power spectrum of the kSZ effect of the Galactic disk found by Ref.~\cite{Hajian:2007xi}.
 
Having developed analytic intuition for the form of the expected screening field, we will now present several physical models for the anisotropic optical depth.
 \begin{figure}[!htp]
\includegraphics[height=20pc]{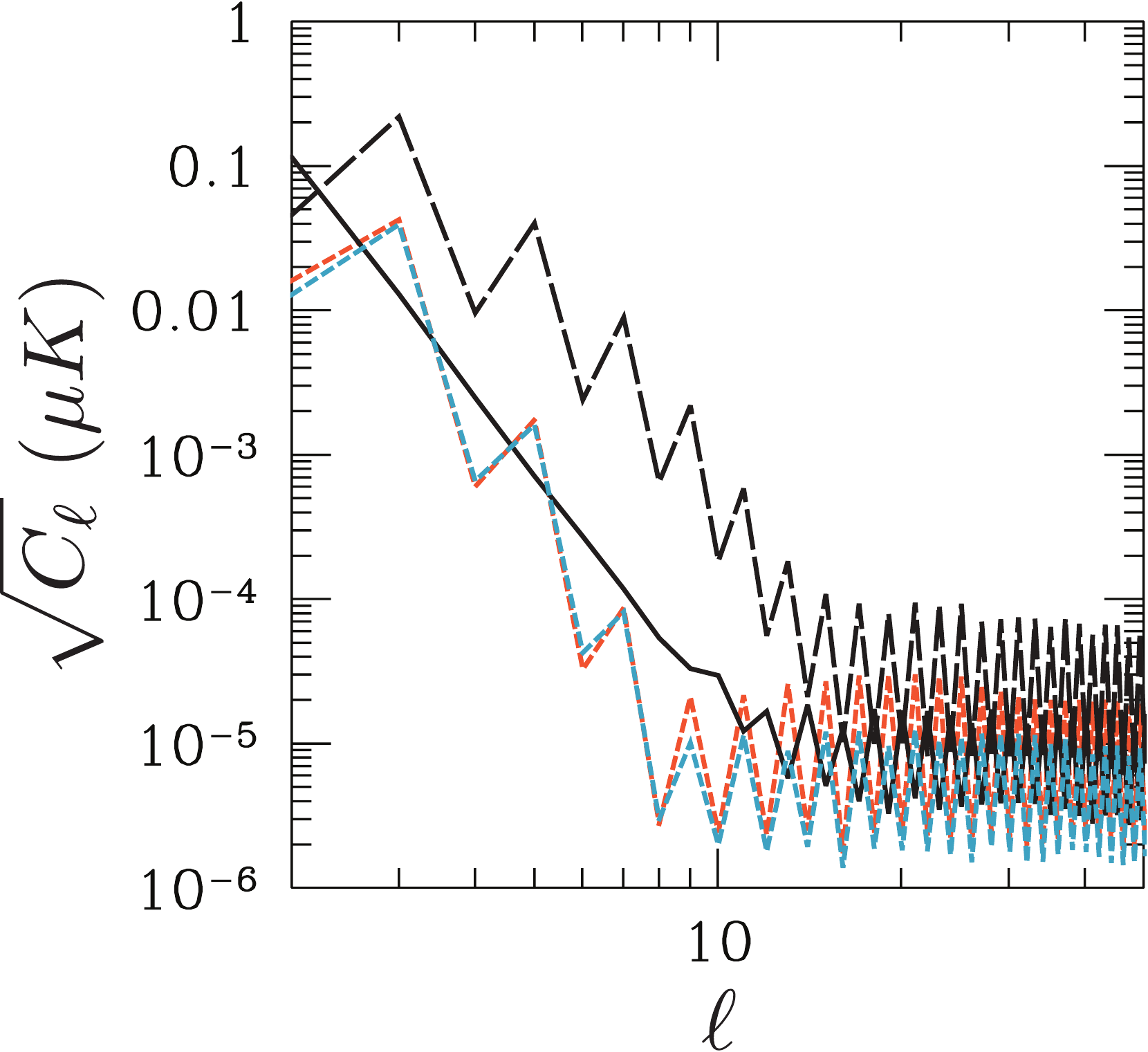}
\caption{The power spectrum of the full screening field corresponding the four models for the free-electron distribution in the halo. The solid black curve corresponds to the generalized NFW halo with a core radius of $r_0 = 300$ kpc  ({\sf SphSym}).  The short-dashed blue and red curves correspond to a triaxial halo (with $e_b = e_c = 0.5$) aligned with ({\sf Triax}) and perpendicular to ({\sf TriaxRot}) the angular momentum of the disk, respectively. The long-dashed black curve corresponds to the triaxial model for the Milky Way halo proposed in Ref.~\cite{Law:2009yq} in order to explain the tidal tails of the Sgr dSph ({\sf TriaxLMJ}).  As described in Eq.~(\ref{eq:angle_avg}) the amplitude of the signal depends linearly on the angle averaged column-density and is shown here for the fiducial case $\bar{\mathcal{C}} = 10^{21}{\rm cm}^{-2}$.  The discussion in Sec.~\ref{sec:doppler_dip} gives an analytic explanation for the sawtooth pattern found in power-spectra generated by these free-electron distributions. }
\label{fig:alms}
\end{figure}

\begin{figure}[!ht]
\includegraphics[height=41pc]{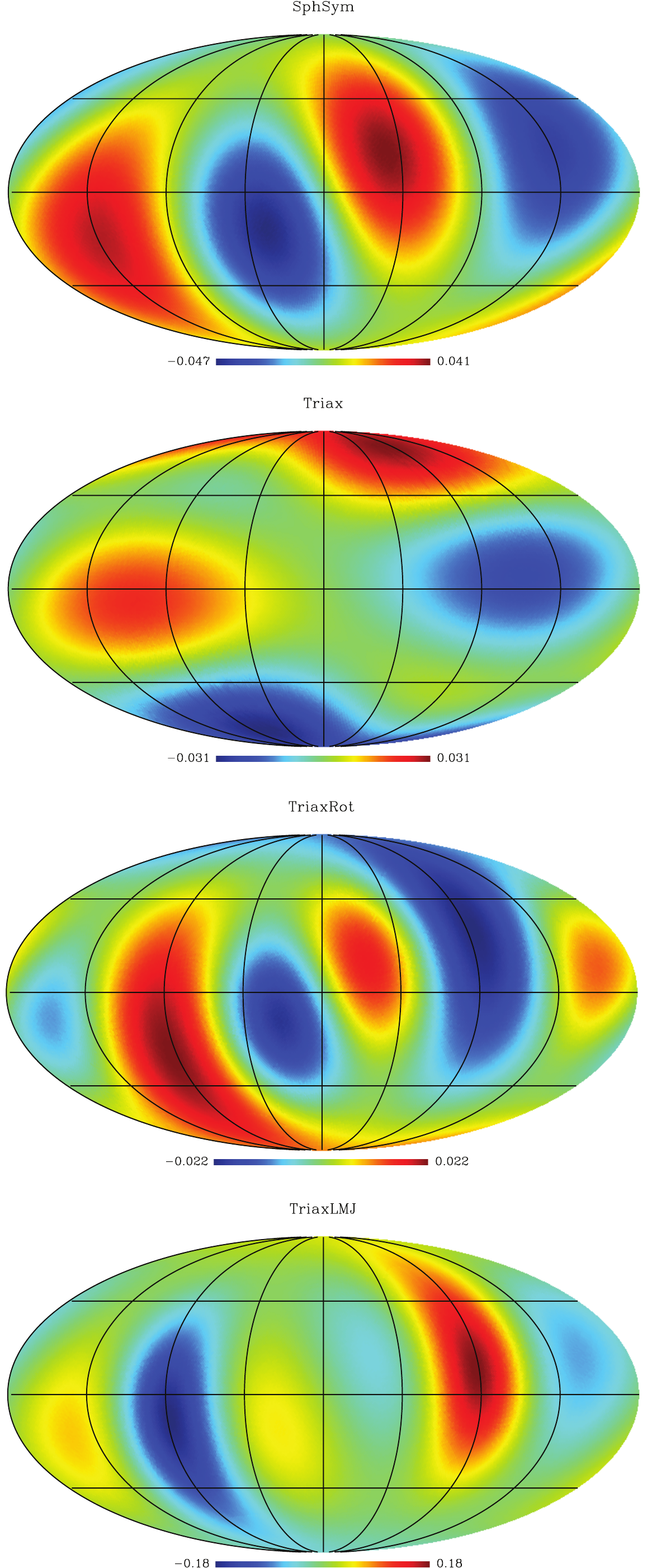}
\caption{Sky maps of the screening field for the three cases, from top to bottom: a generalized spherical NFW halo with $a= 0.01$ ({\sf SphSym}); a generalized triaxial NFW halo with major axis perpendicular to the plane of the Galactic disk, $a=0.01$, and $e_b=e_c =0.5$ ({\sf Triax}); the same as before but with the major axis within the plane of the Galactic disk ({\sf TriaxRot}).  In both cases the intermediate axis is chosen to be perpendicular to the line connecting the Solar system and the Galactic center.  In the final case ({\sf TriaxLMJ}) we take the model for the Galactic halo recently proposed in Ref.~\cite{Law:2009yq} in which the minor axis is aligned with the line connecting the Solar system and Galactic center and the major axis lies within the Galactic disk.}
\label{fig:maps}
\end{figure}
 
\subsection{Physical models for the anisotropic optical depth} \label{sec:anisomodel}

There are two basic ways that we may imagine an anisotropic optical depth through the Galaxy. In the first case, the Galaxy is assumed to have a spherically symmetric gaseous halo and the anisotropy is a consequence of the fact that the Solar system is offset from the Galactic  center. The IAU standard for the Galactocentric distance of the Sun is 8 kpc; however, this value has been recently revised upwards by $\sim 5\%$ \cite{Ghez08, Gillessen09}. We adopt the conservative Galactocentric distance 8.5 kpc to maximize the anisotropic optical depth arising from this offset. In the second case, it is also possible that the gaseous halo is triaxial, leading to an additional source of anisotropy.  We take the virial radius of the Milky Way to be 300 kpc (which corresponds to an NFW halo with a mass of 1.5 $\times 10^{12} \ M_{\odot}$ and a concentration $c=12$).   

\subsubsection{Spherically symmetric halo}

In order to explore the case of a spherically symmetric halo we need only specify the number density of electrons as a function of the radial distance from the Galactic center (GC).  As discussed in Ref.~\cite{Birnboim:2009ne} N-body simulations which include gas dynamics generically predict a hot gaseous halo for Milky Way-like galaxies.  For the spherically symmetric case we will consider a generalized NFW density profile with an inner core
\begin{equation}
n_e = \frac{n_0}{(1+r/r_0)^3}.
\end{equation}
Spherically symmetric profiles can be written in a self-similar form (where the origin is at the GC)
\begin{equation}
n_e = n_0 y_{\rm gas}(r/r_0).
\end{equation}
Using the Galactic coordinate system (where the origin coincides with the location of the Solar system and the GC lies on the positive $x$-axis) the line-of-sight density takes the form
\begin{equation}
n_e(t,\theta,\phi) = n_0 y_{\rm gas}\left(\sqrt{t^2 + a^2 - 2a t \cos(\phi) \sin(\theta)}\right),
\end{equation}
where $t \equiv r/r_0$ and $a \equiv r_{GC}/r_0$.  The normalization of the density distribution is set by requiring the mean angular column density be equal to $10^{21}\ {\rm cm}^{-2}$.  
The normalization for this model can be computed analytically as
\begin{equation}
\bar{\mathcal{C}} = n_0 r_0 \frac{1-a^2 + 2 a^2 \log(a)}{2(1-a^2)^2}.
\end{equation}

\subsubsection{Triaxial halos}

N-body simulations show that dark matter halos are not spherically symmetric but instead are better approximated by triaxial density profiles \cite{1988ApJ...327..507F,1991ApJ...378..496D,1992ApJ...399..405W,Jing:2002np}.  This triaxiality, if reflected in the density distribution of an extended gaseous halo, will produce additional anisotropy in the optical depth.  Fits to N-body simulations \cite{Jing:2002np} show that isodensity contours are well approximated by a radial coordinate
\begin{equation}
R\equiv x^2 + \frac{y^2}{1-e_b^2}+\frac{z^2}{1-e_c^2}, 
\end{equation}
where the $x$-axis and $z$-axis run along the major and minor principal axes, respectively; $e_b$ and $e_c$ ($e_b<e_c$) are the ellipticities of the halo isodensity contours.  Cosmological N-body simulations give a nearly Gaussian distribution for $e_b$ with $\langle e_c \rangle \approx 0.8$ and $e_b \lesssim 0.7$ \cite{Jing:2002np}.  We also note that it has been suggested the inclusion of gas cooling may make the dark-matter density distribution more spherical even out to the virial radius \cite{Kazantzidis:2004vu}.   

As with the spherically symmetric case, we consider a generalized triaxial NFW halo with a core.  As suggested by Ref.~\cite{Zentner:2005wh} in order to obtain consistency between observations of the distribution of sub-halos about the Milky Way and the $\Lambda$CDM prediction, the major axis of the triaxial halo must be aligned with the disk angular momentum.  In order to explore how the relationship of the disk angular momentum and the triaxiality of the dark matter halo affects the local kSZ signal, we consider the two cases where the major axis is aligned with or perpendicular to the disk angular momentum.  In both cases, the intermediate axis is along the Galactic $y$-axis\footnote{The Galactic coordinate system is centered on the location of the Solar system with the Galactic $x$-axis pointing towards the Galactic center and the Galactic $z$-axis pointing in the direction of the angular momentum of the Galactic disk.}.

We also consider the recent triaxial Milky Way halo proposed in Ref.~\cite{Law:2009yq} in order to explain the observed characteristics of the Sagittarius dwarf spheriodal (Sgr dSph).  Using observations of the tidal streams of the Sgr dSph, Ref.~\cite{Law:2009yq} claim a Milky Way halo with a minor/major axis ratio $e_b \approx 0.56$ and an intermediate/major axis ratio of $e_c \approx 0.74$, with the minor axis lying along the Galactic $x$-axis and the major axis along the Galactic $y$-axis.  

\section{Effect of halo kSZ on isotropy anomalies} \label{sec:stats}

Given a model of the anisotropic optical depth, we obtain a prediction for $T_{\rm scr}$ from Eqs.~(\ref{eq:couple}) or ~(\ref{eq:tscreen}). We can add that to the observed realization to obtain the actual cosmological realization, and calculate whatever anomaly statistics we want for the cosmological realization.

\subsection{Quadrupole--octopole planarity and alignment}

First, let us consider the angular momentum dispersion statistics of \S~\ref{sec:impact}. Taking the four models for the halo described in \S~\ref{sec:anisomodel}, labelled {\sf SphSym}, {\sf Triax}, {\sf TriaxRot} and {\sf TriaxLMJ} respectively, we compute the screening fields and add them to the WMAP ILC7 map after subtraction of the kinematic quadrupole. In order to estimate the amplitude of the signal at which the $p$-value of the statistic would be affected for each case, we scale up the maximum of the screening map to an amplitude (1 $\mu$K, 2 $\mu$K, 5 $\mu$K) which is much smaller than the primordial quadrupole and octopole signals, before adding to the ILC7 map. The 1 $\mu K$ signal is not large enough to make any difference to the statistics. The results for the other two cases are presented for the $L^2_{23}$ statistic in Table~\ref{tab:screeningstats}. The $L^2_2$ and $L^2_3$ statistics of the ILC7-KQ map are not anomalous before or after the kSZ correction and thus we do not repeat them here. 

\begin{table}[!ht]
\caption{Values and $p$-values (computed from 10,000 isotropic realizations) for angular momentum dispersion statistic $L^2_{23}$ for the WMAP 7 year Internal Linear Combination (ILC7) map subtracting the kinematic quadrupole (KQ) after accounting for the anisotropic screening from the four models described in the text, after scaling the maximum amplitude of the screening map to 2 $\mu$K and 5 $\mu$K.  We also list the corresponding value for the angle-averaged free electron column density, $\bar{\mathcal{C}}$, in units of $10^{21}\ {\rm cm}^{-2}$.}
\begin{center}
\begin{tabular}{| l | c|c|c||c|c|c|}
\hline
Model & $L^2_{23}$  & $p$-value & $\bar{\mathcal{C}}$ & $L^2_{23}$ & $p$-value & $\bar{\mathcal{C}}$\\
	& (2 $\mu$K) &   (2 $\mu$K) & (2 $\mu$K) & (5 $\mu$K) & (5 $\mu$K) & (5 $\mu$K)  \\
\hline
{\sf SphSym} &	 0.955	&	0.0047	&	12 	&	0.952	&  0.0056 	&	31 \\
{\sf Triax} 	& 0.962	& 	0.0022	&	22 	&  	0.972	&    0.0008	&	56 \\
{\sf TriaxRot} & 	 0.950	&  	0.0062	&	24 	& 	0.943	&  0.0091 	&	60 \\
{\sf TriaxLMJ}  &  0.956	& 	0.0042	&	4 	& 	0.958	& 	0.0036	&	11 	\\
\hline
\end{tabular}
\end{center}
\label{tab:screeningstats}
\end{table}

In interpreting Table~\ref{tab:screeningstats}, the baseline $p$-value to compare with is $0.0055$ for the ILC7-KQ map with no kSZ correction. The Table shows that for the {\sf Triax} case, small signals at the $\sim 2-5$ $\mu$K level {\sl reduce} this already small $p$-value by factors of $\sim 2$--$7$! When scaled to $5$ $\mu$K, the {\sf TriaxRot} $p$-value increases somewhat by factors of $\lesssim 2$ and the {\sf TriaxLMJ} $p$-value decreases by roughly the same factor. The {\sf SphSym} screening map is unable to affect this statistic significantly at these signal levels. These tests show that a statistic such as $L^2_{23}$, which is sensitive to the orientation of the Galactic plane,  can be highly sensitive to small signals oriented with this plane. 

If the maximum amplitude of the screening map is between 2--5 $\mu$K, the average free electron column densities for these maps are between $4 \times 10^{21}$ --  $6 \times 10^{22}\ {\rm cm}^{-2}$.  As we discuss in more detail below, though the lower end of this range is possible given current observational constraints, the upper range is unlikely.  

\subsection{Hemispherical asymmetry}
 
We now explore the extent to which the scattering of CMB photons in an extended gaseous halo can account for the observed hemispherical asymmetry \cite{Eriksen:2003db}. The statistic by which this asymmetry is inferred assumes a dipolar modulation of an isotropic primordial CMB temperature field, and constrains the relative amplitude of this modulation. The dipolar modulation leads to a coupling between modes of order $\ell$ and $\ell\pm1$.  As we saw in Sec.~\ref{sec:doppler_dip}, the kSZ also introduces such a coupling between modes, so it is particularly interesting to consider how, through the kSZ, a screening field may impact the significance of the dipole modulation amplitude statistic for constraining hemispherical asymmetry. 
 
In order to do this, we will use the approach presented in Refs.~\cite{Gordon:2005ai, Dvorkin:2007jp}.  In that work an estimator is derived which measures to what extent the data shows a dipolar modulation by looking for a non-zero coupling between multipoles of order $\ell$ and $\ell\pm1$ up to some $\ell_{\rm max}$, 
 \begin{equation}
 \hat{w}_1^{TT} = \frac{\sum_{\ell m} \frac{f_{\ell}^{TT} R^{1 \ell}_{\ell+1,m}}{C_{\ell}^{TT} C_{\ell+1}^{TT}}(a^T_{\ell m})^* (a^T_{\ell+1 m})}{\sum_{\ell m} \frac{(f_{\ell}^{TT} R^{1 \ell}_{\ell+1,m})^2}{C^{TT}_{\ell} C^{TT}_{\ell+1 m}}},
 \end{equation}
 where 
 \begin{eqnarray}
 f_{\ell}^{TT} &\equiv& C_{\ell}^{TT} + C_{\ell+1}^{TT}\\
 R^{\ell_1, \ell_2}_{\ell m} &\equiv& (-1)^m \sqrt{\frac{(2 \ell+1)(2 \ell_1+1)(2 \ell_2 +1)}{4\pi}} \times \nonumber \\
 & & \left(\begin{array}{ccc}\ell_1 & \ell_2 & \ell \\0 & 0 & 0\end{array}\right)\left(\begin{array}{ccc}\ell_1 & \ell_2 & \ell \\0 & m & -m\end{array}\right).
 \end{eqnarray}
 From the discussion in Sec.~\ref{sec:doppler_dip} it is clear that the inclusion of Thomson scattering from free electrons in an extended Galactic halo will couple multipoles of order $\ell$ to those of order $\ell-1$ and $\ell +1$ in the observed $a_{\ell m}$s.  

\begin{figure}[!ht]
\includegraphics[height=20pc]{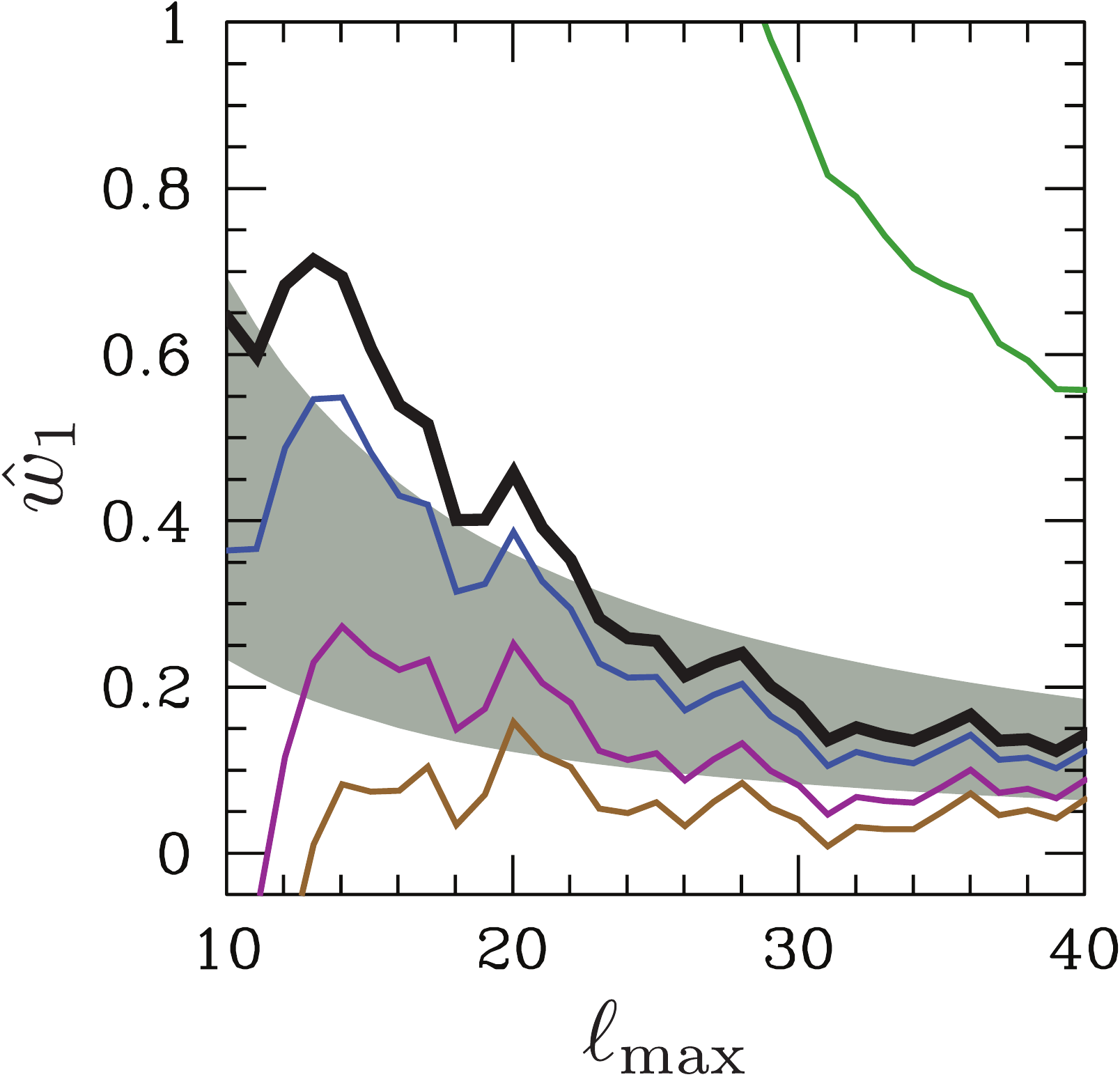}
\caption{How a screening field can affect the measurement of a hemispherical asymmetry.  The grey band indicates the confidence level of the distribution at that value of $\ell_{\rm max}$ ranging from 1 $\sigma$ to 3 $\sigma$.  The solid thick black curve shows the value of $\hat{w}_1$ evaluated on the ILC cleaned WMAP7 map.  From top to bottom we show the resulting hemispherical asymmetry for {\sf TriaxLMJ}, {\sf SphSym}, {\sf TriaxRot}, and {\sf Triax} as a function of $\ell_{\rm max}$.  This Figure demonstrates how the orientation of the triaxiality of the halo greatly affects its ability to impact the inferred amplitude of a hemispherical asymmetry.}
\label{fig:hemiasym}
\end{figure}
 
To check the effect of the kSZ on the significance of the hemispherical asymmetry, we evaluate the estimator $\hat{w}_1^{TT}$ on the inferred cosmological signal given by $T^{\rm cos}_{\ell m} = T^{\rm obs}_{\ell m} +  T^{\rm scr}_{\ell m}(\bar{\mathcal{C}})$ for various values of the angle-averaged electron column density $\bar{\mathcal{C}}$.  As in the previous section, the four extended halos considered here are the spherically symmetric case ({\sf SphSym}), and the three triaxial halos with various orientations ({\sf Triax}, {\sf TriaxRot}, {\sf TriaxLMJ}).  

We show the value of the estimator $\hat{w}_1$ evaluated on the WMAP ILC7 map in Fig.~\ref{fig:hemiasym}.  Because of the scale dependence of the power spectra of the screening fields, seen in Fig.~\ref{fig:alms}, we plot the value of $\hat{w}_1$ as a function of $\ell_{\rm max}$ in Fig.~\ref{fig:hemiasym}.  
The grey band indicates the confidence level of the distribution at that value of $\ell_{\rm max}$ ranging from 1 $\sigma$ to 3 $\sigma$.  The Figure shows the value of $\hat{w}_1$ for (from top to bottom) for {\sf TriaxLMJ}, {\sf SphSym}, {\sf TriaxRot}, and {\sf Triax}.  It is clear that the orientation of any triaxiality has a significant impact on the inferred amplitude of the hemispherical asymmetry.  In particular, the {\sf TriaxLMJ} orientation leads to an increase in the inferred amplitude.  This can be seen in Fig.~\ref{fig:maps} since the inclusion of the {\sf TriaxLMJ} screening field adds power in the northern Galactic hemisphere and subtracts it from the southern hemisphere. For the other cases, the inferred amplitude decreases.

Since the anisotropies induced by a local kSZ are at the level of $10^{-5}\ \mu{\rm K}$ for $\ell \gtrsim 10$ (see Fig.~\ref{fig:alms}), in order for the screening field to have a significant impact on the inferred amplitude of $\hat{w}_1$, we must have $\bar{\mathcal{C}}/(10^{21}\ {\rm cm^{-2}}) \sim 10^4$.  A free-electron column density of this magnitude is ruled out by observations of OVII and OVIII absorption \cite{2007ApJ...669..990B} and pulsar observations towards the Large Magellanic Cloud (LMC) \cite{Taylor:1993my}.  Therefore, we find that a local kSZ is unlikely to provide a plausible explanation for a hemispherical asymmetry.

\section{Discussion} \label{sec:discuss}

Although measurements of the CMB have generally confirmed our current understanding of the formation and evolution of the universe, there are several anomalies which appear to be at odds with our standard picture.
These anomalies could suggest that there may be some process which violates the statistical isotropy of the CMB fluctuations on large scales.  Although attempts have been made to modify the physics of the early universe in order to explain these anomalies, there are also strong reasons to look for explanations in the local universe.  First, the directionality of these anomalies seems to closely coincide with both the Solar CMB dipole as well as the velocity of the Galactic barycenter relative to the CMB rest-frame.  Second, given that the anomalies are on large angular scales, any non-primordial and causal explanation must be local in origin.  

Here we have explored how a local kSZ signal (due to the motion of any extended hot gaseous halo associated with the Milky Way relative to the CMB) may affect the significance of several anomalies observed in the CMB anisotropies.  Both theoretical and observational considerations indicate the presence of a hot gaseous halo with an extent of several tens of kiloparsecs \cite{Bregman:2009qh, Birnboim:2009ne}.  Anisotropies in the optical depth through this gaseous halo can be due to the offset of the Solar system from the Galactic center, as well as any triaxiality in its distribution. We computed the kSZ signal from several plausible physical models for the shape and orientation of the halo, and studied their impact on the observed CMB sky, in the form of the WMAP ILC7 map.

We considered how a local kSZ may affect the observed planarity of the CMB quadrupole and octopole moments as well as their relative alignment, motivated by the fact that kSZ from a triaxial halo can naturally relate to the special directions that appear to be associated with the $\ell=2$ and $3$ multipole moments. Surprisingly, we found that relatively small changes in the observed amplitude of these moments ($\sim 10\%$) to account for a kSZ signal can reduce the already tiny $p$-values for these anomalies of up to factors of 2--7.  The corresponding free electron column density needed to affect the planarity/alignment statistics at this level is $4$ -- $60 \times 10^{21}\ {\rm cm}^{-2}$. 

A local kSZ signal couples multipole moments of order $\ell$ to those of order $\ell \pm 1$, so it is natural to consider how such a signal would affect any inferred hemispherical asymmetry in the CMB anisotropies in the form of a dipolar modulation, which gives rise to the same coupling.  Using a statistic first derived in Ref.~\cite{Dvorkin:2007jp}, we found that a local kSZ signal can have a significant effect on the inferred amplitude of a dipolar modulation of the primary CMB anisotropies.  In this case, the corresponding free electron column density would need to be $\sim10^{25}\ {\rm cm}^{-2}$.

Theoretical and observational constraints on the free electron fraction in a extended hot gaseous halo associated with the Milky Way place a fairly strict bound on the column density of $< 10^{21}\ {\rm cm}^{-2}$.  The most precise constraints come from observations of the dispersion measure to individual pulsars \cite{Taylor:1993my} which find a column density in free electrons to the LMC of $\sim 3 \times 10^{20}\ {\rm cm}^{-2}$.  Given that the LMC is $\sim 50$ kpc from the Galactic center and that the scale-radius of the Milky Way halo is $\sim 300$ kpc, these observations indicate that the total optical depth may be as large as $10^{21}\ {\rm cm}^{-2}$.  In addition to this, observations of local ($z=0$) OVII and OVIII absorption towards several quasars indicates the presence of an extended hot gaseous halo around the Milky Way \cite{Bregman:2009qh}.  The inferred free electron column density is quite uncertain given assumptions about the metallicity of the gas (Solar abundance was assumed to convert the observations to an electron density; sub-Solar values would increase the inferred density), as well as a model dependence coming from the assumed profile of the gas.  Finally, theoretical considerations imply the existence of a hot extended gaseous halo with a fractional mass of the order of the cosmic baryon fraction, $f\sim 0.1$.  Therefore, for a Milky Way sized halo ($M \sim 1.5 \times 10^{12}\ M_{\odot}$) with a scale radius of 300 kpc, we would expect a column density no larger than $10^{21}\ {\rm cm}^{-2}$. 
Given that, for a local kSZ signal to have a significant effect on the CMB anisotropies, we must have a free electron column density $> 10^{21} \ {\rm cm}^{-2}$, but that both theoretical and observational considerations place the limit at $\lesssim 10^{21} \ {\rm cm}^{-2}$, it is unlikely that a local kSZ signal can explain any of the CMB anomalies considered here. Even given the uncertainties, the kSZ signal can at best only be at the lowest amplitude needed to affect the anomaly statistics.

Finally, we note that contamination from kSZ in the Solar system can also, in principle, provide an anisotropic contamination of the CMB sky at large angles. The geometry of the heliopause is coincidentally aligned with the CMB dipole \cite{Frisch:2007kz}, providing a motivation for looking there for a explanation for the special directional properties of the large-angle isotropy anomalies. However, in practice, the optical depth of free electrons in the Solar system is more than 7 orders of magnitude below what is required to produce the necessary signal.

\acknowledgments{HVP is supported by Marie Curie grant MIRG-CT-2007-203314 from the European Commission, and by STFC and the Leverhulme Trust.  TLS is supported by the Berkeley Center for Cosmological Physics.  We thank the Aspen Center for Physics, where this work was initiated, for hospitality. We are grateful to Hsiao-Wen Chen, Priscilla Frisch, Shirley Ho, Ed Jenkins, and Chris Thom for invaluable information about observational constraints on the free electron content of the local universe. We thank Yuval Birnboim for useful comments on some preliminary results.  HVP thanks Anthony Challinor, Daniel Mortlock, and Andrew Pontzen for interesting discussions on related topics.}


\begin{thebibliography}{49}
\expandafter\ifx\csname natexlab\endcsname\relax\def\natexlab#1{#1}\fi
\expandafter\ifx\csname bibnamefont\endcsname\relax
  \def\bibnamefont#1{#1}\fi
\expandafter\ifx\csname bibfnamefont\endcsname\relax
  \def\bibfnamefont#1{#1}\fi
\expandafter\ifx\csname citenamefont\endcsname\relax
  \def\citenamefont#1{#1}\fi
\expandafter\ifx\csname url\endcsname\relax
  \def\url#1{\texttt{#1}}\fi
\expandafter\ifx\csname urlprefix\endcsname\relax\def\urlprefix{URL }\fi
\providecommand{\bibinfo}[2]{#2}
\providecommand{\eprint}[2][]{\url{#2}}

\bibitem[{\citenamefont{Copi et~al.}(2004)\citenamefont{Copi, Huterer, and
  Starkman}}]{Copi:2003kt}
\bibinfo{author}{\bibfnamefont{C.~J.} \bibnamefont{Copi}},
  \bibinfo{author}{\bibfnamefont{D.}~\bibnamefont{Huterer}}, \bibnamefont{and}
  \bibinfo{author}{\bibfnamefont{G.~D.} \bibnamefont{Starkman}},
  \bibinfo{journal}{Phys. Rev.} \textbf{\bibinfo{volume}{D70}},
  \bibinfo{pages}{043515} (\bibinfo{year}{2004}), \eprint{astro-ph/0310511}.

\bibitem[{\citenamefont{de~Oliveira-Costa
  et~al.}(2004)\citenamefont{de~Oliveira-Costa, Tegmark, Zaldarriaga, and
  Hamilton}}]{deOliveiraCosta:2003pu}
\bibinfo{author}{\bibfnamefont{A.}~\bibnamefont{de~Oliveira-Costa}},
  \bibinfo{author}{\bibfnamefont{M.}~\bibnamefont{Tegmark}},
  \bibinfo{author}{\bibfnamefont{M.}~\bibnamefont{Zaldarriaga}},
  \bibnamefont{and} \bibinfo{author}{\bibfnamefont{A.}~\bibnamefont{Hamilton}},
  \bibinfo{journal}{Phys. Rev.} \textbf{\bibinfo{volume}{D69}},
  \bibinfo{pages}{063516} (\bibinfo{year}{2004}), \eprint{astro-ph/0307282}.

\bibitem[{\citenamefont{Eriksen et~al.}(2004)\citenamefont{Eriksen, Hansen,
  Banday, Gorski, and Lilje}}]{Eriksen:2003db}
\bibinfo{author}{\bibfnamefont{H.~K.} \bibnamefont{Eriksen}},
  \bibinfo{author}{\bibfnamefont{F.~K.} \bibnamefont{Hansen}},
  \bibinfo{author}{\bibfnamefont{A.~J.} \bibnamefont{Banday}},
  \bibinfo{author}{\bibfnamefont{K.~M.} \bibnamefont{Gorski}},
  \bibnamefont{and} \bibinfo{author}{\bibfnamefont{P.~B.} \bibnamefont{Lilje}},
  \bibinfo{journal}{Astrophys. J.} \textbf{\bibinfo{volume}{605}},
  \bibinfo{pages}{14} (\bibinfo{year}{2004}), \eprint{astro-ph/0307507}.

\bibitem[{\citenamefont{Schwarz et~al.}(2004)\citenamefont{Schwarz, Starkman,
  Huterer, and Copi}}]{Schwarz:2004gk}
\bibinfo{author}{\bibfnamefont{D.~J.} \bibnamefont{Schwarz}},
  \bibinfo{author}{\bibfnamefont{G.~D.} \bibnamefont{Starkman}},
  \bibinfo{author}{\bibfnamefont{D.}~\bibnamefont{Huterer}}, \bibnamefont{and}
  \bibinfo{author}{\bibfnamefont{C.~J.} \bibnamefont{Copi}},
  \bibinfo{journal}{Phys. Rev. Lett.} \textbf{\bibinfo{volume}{93}},
  \bibinfo{pages}{221301} (\bibinfo{year}{2004}), \eprint{astro-ph/0403353}.

\bibitem[{\citenamefont{Land and Magueijo}(2005)}]{Land:2005ad}
\bibinfo{author}{\bibfnamefont{K.}~\bibnamefont{Land}} \bibnamefont{and}
  \bibinfo{author}{\bibfnamefont{J.}~\bibnamefont{Magueijo}},
  \bibinfo{journal}{Phys. Rev. Lett.} \textbf{\bibinfo{volume}{95}},
  \bibinfo{pages}{071301} (\bibinfo{year}{2005}), \eprint{astro-ph/0502237}.

\bibitem[{\citenamefont{Groeneboom and Eriksen}(2009)}]{Groeneboom:2008fz}
\bibinfo{author}{\bibfnamefont{N.~E.} \bibnamefont{Groeneboom}}
  \bibnamefont{and} \bibinfo{author}{\bibfnamefont{H.~K.}
  \bibnamefont{Eriksen}}, \bibinfo{journal}{Astrophys. J.}
  \textbf{\bibinfo{volume}{690}}, \bibinfo{pages}{1807} (\bibinfo{year}{2009}),
  \eprint{0807.2242}.

\bibitem[{\citenamefont{Hanson and Lewis}(2009)}]{Hanson:2009gu}
\bibinfo{author}{\bibfnamefont{D.}~\bibnamefont{Hanson}} \bibnamefont{and}
  \bibinfo{author}{\bibfnamefont{A.}~\bibnamefont{Lewis}},
  \bibinfo{journal}{Phys. Rev.} \textbf{\bibinfo{volume}{D80}},
  \bibinfo{pages}{063004} (\bibinfo{year}{2009}), \eprint{0908.0963}.

\bibitem[{\citenamefont{Groeneboom et~al.}(2009)\citenamefont{Groeneboom,
  Ackerman, Wehus, and Eriksen}}]{Groeneboom:2009cb}
\bibinfo{author}{\bibfnamefont{N.~E.} \bibnamefont{Groeneboom}},
  \bibinfo{author}{\bibfnamefont{L.}~\bibnamefont{Ackerman}},
  \bibinfo{author}{\bibfnamefont{I.~K.} \bibnamefont{Wehus}}, \bibnamefont{and}
  \bibinfo{author}{\bibfnamefont{H.~K.} \bibnamefont{Eriksen}}
  (\bibinfo{year}{2009}), \eprint{0911.0150}.

\bibitem[{\citenamefont{Land and Magueijo}(2007)}]{Land:2006bn}
\bibinfo{author}{\bibfnamefont{K.}~\bibnamefont{Land}} \bibnamefont{and}
  \bibinfo{author}{\bibfnamefont{J.}~\bibnamefont{Magueijo}},
  \bibinfo{journal}{Mon. Not. Roy. Astron. Soc.}
  \textbf{\bibinfo{volume}{378}}, \bibinfo{pages}{153} (\bibinfo{year}{2007}),
  \eprint{astro-ph/0611518}.

\bibitem[{\citenamefont{Hansen et~al.}(2009)\citenamefont{Hansen, Banday,
  Gorski, Eriksen, and Lilje}}]{Hansen:2008ym}
\bibinfo{author}{\bibfnamefont{F.~K.} \bibnamefont{Hansen}},
  \bibinfo{author}{\bibfnamefont{A.~J.} \bibnamefont{Banday}},
  \bibinfo{author}{\bibfnamefont{K.~M.} \bibnamefont{Gorski}},
  \bibinfo{author}{\bibfnamefont{H.~K.} \bibnamefont{Eriksen}},
  \bibnamefont{and} \bibinfo{author}{\bibfnamefont{P.~B.} \bibnamefont{Lilje}},
  \bibinfo{journal}{Astrophys. J.} \textbf{\bibinfo{volume}{704}},
  \bibinfo{pages}{1448} (\bibinfo{year}{2009}), \eprint{0812.3795}.

\bibitem[{\citenamefont{Ackerman et~al.}(2007)\citenamefont{Ackerman, Carroll,
  and Wise}}]{Ackerman:2007nb}
\bibinfo{author}{\bibfnamefont{L.}~\bibnamefont{Ackerman}},
  \bibinfo{author}{\bibfnamefont{S.~M.} \bibnamefont{Carroll}},
  \bibnamefont{and} \bibinfo{author}{\bibfnamefont{M.~B.} \bibnamefont{Wise}},
  \bibinfo{journal}{Phys. Rev.} \textbf{\bibinfo{volume}{D75}},
  \bibinfo{pages}{083502} (\bibinfo{year}{2007}), \eprint{astro-ph/0701357}.

\bibitem[{\citenamefont{Erickcek
  et~al.}(2008{\natexlab{a}})\citenamefont{Erickcek, Kamionkowski, and
  Carroll}}]{Erickcek:2008sm}
\bibinfo{author}{\bibfnamefont{A.~L.} \bibnamefont{Erickcek}},
  \bibinfo{author}{\bibfnamefont{M.}~\bibnamefont{Kamionkowski}},
  \bibnamefont{and} \bibinfo{author}{\bibfnamefont{S.~M.}
  \bibnamefont{Carroll}}, \bibinfo{journal}{Phys. Rev.}
  \textbf{\bibinfo{volume}{D78}}, \bibinfo{pages}{123520}
  (\bibinfo{year}{2008}{\natexlab{a}}), \eprint{0806.0377}.

\bibitem[{\citenamefont{Erickcek
  et~al.}(2008{\natexlab{b}})\citenamefont{Erickcek, Carroll, and
  Kamionkowski}}]{Erickcek:2008jp}
\bibinfo{author}{\bibfnamefont{A.~L.} \bibnamefont{Erickcek}},
  \bibinfo{author}{\bibfnamefont{S.~M.} \bibnamefont{Carroll}},
  \bibnamefont{and}
  \bibinfo{author}{\bibfnamefont{M.}~\bibnamefont{Kamionkowski}},
  \bibinfo{journal}{Phys. Rev.} \textbf{\bibinfo{volume}{D78}},
  \bibinfo{pages}{083012} (\bibinfo{year}{2008}{\natexlab{b}}),
  \eprint{0808.1570}.

\bibitem[{\citenamefont{Erickcek et~al.}(2009)\citenamefont{Erickcek, Hirata,
  and Kamionkowski}}]{Erickcek:2009at}
\bibinfo{author}{\bibfnamefont{A.~L.} \bibnamefont{Erickcek}},
  \bibinfo{author}{\bibfnamefont{C.~M.} \bibnamefont{Hirata}},
  \bibnamefont{and}
  \bibinfo{author}{\bibfnamefont{M.}~\bibnamefont{Kamionkowski}},
  \bibinfo{journal}{Phys. Rev.} \textbf{\bibinfo{volume}{D80}},
  \bibinfo{pages}{083507} (\bibinfo{year}{2009}), \eprint{0907.0705}.

\bibitem[{\citenamefont{Dvorkin et~al.}(2008)\citenamefont{Dvorkin, Peiris, and
  Hu}}]{Dvorkin:2007jp}
\bibinfo{author}{\bibfnamefont{C.}~\bibnamefont{Dvorkin}},
  \bibinfo{author}{\bibfnamefont{H.~V.} \bibnamefont{Peiris}},
  \bibnamefont{and} \bibinfo{author}{\bibfnamefont{W.}~\bibnamefont{Hu}},
  \bibinfo{journal}{Phys. Rev.} \textbf{\bibinfo{volume}{D77}},
  \bibinfo{pages}{063008} (\bibinfo{year}{2008}), \eprint{0711.2321}.

\bibitem[{\citenamefont{Hirata}(2009)}]{Hirata:2009ar}
\bibinfo{author}{\bibfnamefont{C.~M.} \bibnamefont{Hirata}},
  \bibinfo{journal}{JCAP} \textbf{\bibinfo{volume}{0909}}, \bibinfo{pages}{011}
  (\bibinfo{year}{2009}), \eprint{0907.0703}.

\bibitem[{\citenamefont{Frommert and Ensslin}(2009)}]{Frommert:2009qw}
\bibinfo{author}{\bibfnamefont{M.}~\bibnamefont{Frommert}} \bibnamefont{and}
  \bibinfo{author}{\bibfnamefont{T.~A.} \bibnamefont{Ensslin}}
  (\bibinfo{year}{2009}), \eprint{0908.0453}.

\bibitem[{\citenamefont{Frisch}(2007)}]{Frisch:2007kz}
\bibinfo{author}{\bibfnamefont{P.~C.} \bibnamefont{Frisch}}
  (\bibinfo{year}{2007}), \eprint{0707.2970}.

\bibitem[{\citenamefont{{Hajian} et~al.}(2007)\citenamefont{{Hajian},
  {Hern{\'a}ndez-Monteagudo}, {Jimenez}, {Spergel}, and
  {Verde}}}]{Hajian:2007xi}
\bibinfo{author}{\bibfnamefont{A.}~\bibnamefont{{Hajian}}},
  \bibinfo{author}{\bibfnamefont{C.}~\bibnamefont{{Hern{\'a}ndez-Monteagudo}}},
  \bibinfo{author}{\bibfnamefont{R.}~\bibnamefont{{Jimenez}}},
  \bibinfo{author}{\bibfnamefont{D.}~\bibnamefont{{Spergel}}},
  \bibnamefont{and} \bibinfo{author}{\bibfnamefont{L.}~\bibnamefont{{Verde}}},
  \bibinfo{journal}{\apj} \textbf{\bibinfo{volume}{671}}, \bibinfo{pages}{1079}
  (\bibinfo{year}{2007}), \eprint{0705.3245}.

\bibitem[{\citenamefont{{Waelkens} et~al.}(2008)\citenamefont{{Waelkens},
  {Maturi}, and {En{\ss}lin}}}]{Waelkens:2007wn}
\bibinfo{author}{\bibfnamefont{A.}~\bibnamefont{{Waelkens}}},
  \bibinfo{author}{\bibfnamefont{M.}~\bibnamefont{{Maturi}}}, \bibnamefont{and}
  \bibinfo{author}{\bibfnamefont{T.}~\bibnamefont{{En{\ss}lin}}},
  \bibinfo{journal}{\mnras} \textbf{\bibinfo{volume}{383}},
  \bibinfo{pages}{1425} (\bibinfo{year}{2008}), \eprint{0707.2601}.

\bibitem[{\citenamefont{{Suto} et~al.}(1996)\citenamefont{{Suto}, {Makishima},
  {Ishisaki}, and {Ogasaka}}}]{Suto1996}
\bibinfo{author}{\bibfnamefont{Y.}~\bibnamefont{{Suto}}},
  \bibinfo{author}{\bibfnamefont{K.}~\bibnamefont{{Makishima}}},
  \bibinfo{author}{\bibfnamefont{Y.}~\bibnamefont{{Ishisaki}}},
  \bibnamefont{and}
  \bibinfo{author}{\bibfnamefont{Y.}~\bibnamefont{{Ogasaka}}},
  \bibinfo{journal}{\apjl} \textbf{\bibinfo{volume}{461}},
  \bibinfo{pages}{L33+} (\bibinfo{year}{1996}),
  \eprint{arXiv:astro-ph/9602061}.

\bibitem[{\citenamefont{Abramo et~al.}(2006)\citenamefont{Abramo, Jr., and
  Wuensche}}]{Abramo:2006hs}
\bibinfo{author}{\bibfnamefont{L.~R.} \bibnamefont{Abramo}},
  \bibinfo{author}{\bibfnamefont{L.~S.} \bibnamefont{Jr.}}, \bibnamefont{and}
  \bibinfo{author}{\bibfnamefont{C.~A.} \bibnamefont{Wuensche}},
  \bibinfo{journal}{Phys. Rev.} \textbf{\bibinfo{volume}{D74}},
  \bibinfo{pages}{083515} (\bibinfo{year}{2006}), \eprint{astro-ph/0605269}.

\bibitem[{\citenamefont{{Maturi} et~al.}(2007)\citenamefont{{Maturi}, {Dolag},
  {Waelkens}, {Springel}, and {En{\ss}lin}}}]{Maturi:2007xr}
\bibinfo{author}{\bibfnamefont{M.}~\bibnamefont{{Maturi}}},
  \bibinfo{author}{\bibfnamefont{K.}~\bibnamefont{{Dolag}}},
  \bibinfo{author}{\bibfnamefont{A.}~\bibnamefont{{Waelkens}}},
  \bibinfo{author}{\bibfnamefont{V.}~\bibnamefont{{Springel}}},
  \bibnamefont{and}
  \bibinfo{author}{\bibfnamefont{T.}~\bibnamefont{{En{\ss}lin}}},
  \bibinfo{journal}{Astron. Astrophys.} \textbf{\bibinfo{volume}{476}},
  \bibinfo{pages}{83} (\bibinfo{year}{2007}), \eprint{0708.1881}.

\bibitem[{\citenamefont{Francis and Peacock}(2009)}]{Francis:2009pt}
\bibinfo{author}{\bibfnamefont{C.~L.} \bibnamefont{Francis}} \bibnamefont{and}
  \bibinfo{author}{\bibfnamefont{J.~A.} \bibnamefont{Peacock}}
  (\bibinfo{year}{2009}), \eprint{0909.2495}.

\bibitem[{\citenamefont{Birnboim and Loeb}(2009)}]{Birnboim:2009ne}
\bibinfo{author}{\bibfnamefont{Y.}~\bibnamefont{Birnboim}} \bibnamefont{and}
  \bibinfo{author}{\bibfnamefont{A.}~\bibnamefont{Loeb}},
  \bibinfo{journal}{JCAP} \textbf{\bibinfo{volume}{0906}}, \bibinfo{pages}{008}
  (\bibinfo{year}{2009}), \eprint{0903.3943}.

\bibitem[{\citenamefont{Keres et~al.}(2005)\citenamefont{Keres, Katz, Weinberg,
  and Dave}}]{Keres:2004cq}
\bibinfo{author}{\bibfnamefont{D.}~\bibnamefont{Keres}},
  \bibinfo{author}{\bibfnamefont{N.}~\bibnamefont{Katz}},
  \bibinfo{author}{\bibfnamefont{D.~H.} \bibnamefont{Weinberg}},
  \bibnamefont{and} \bibinfo{author}{\bibfnamefont{R.}~\bibnamefont{Dave}},
  \bibinfo{journal}{Mon. Not. Roy. Astron. Soc.}
  \textbf{\bibinfo{volume}{363}}, \bibinfo{pages}{2} (\bibinfo{year}{2005}),
  \eprint{astro-ph/0407095}.

\bibitem[{\citenamefont{Birnboim and Dekel}(2003)}]{Birnboim:2003xa}
\bibinfo{author}{\bibfnamefont{Y.}~\bibnamefont{Birnboim}} \bibnamefont{and}
  \bibinfo{author}{\bibfnamefont{A.}~\bibnamefont{Dekel}},
  \bibinfo{journal}{Mon. Not. Roy. Astron. Soc.}
  \textbf{\bibinfo{volume}{345}}, \bibinfo{pages}{349} (\bibinfo{year}{2003}),
  \eprint{astro-ph/0302161}.

\bibitem[{\citenamefont{{Bregman} and
  {Lloyd-Davies}}(2007)}]{2007ApJ...669..990B}
\bibinfo{author}{\bibfnamefont{J.~N.} \bibnamefont{{Bregman}}}
  \bibnamefont{and} \bibinfo{author}{\bibfnamefont{E.~J.}
  \bibnamefont{{Lloyd-Davies}}}, \bibinfo{journal}{\apj}
  \textbf{\bibinfo{volume}{669}}, \bibinfo{pages}{990} (\bibinfo{year}{2007}),
  \eprint{0707.1699}.

\bibitem[{\citenamefont{Taylor and Cordes}(1993)}]{Taylor:1993my}
\bibinfo{author}{\bibfnamefont{J.~H.} \bibnamefont{Taylor}} \bibnamefont{and}
  \bibinfo{author}{\bibfnamefont{J.~M.} \bibnamefont{Cordes}},
  \bibinfo{journal}{Astrophys. J.} \textbf{\bibinfo{volume}{411}},
  \bibinfo{pages}{674} (\bibinfo{year}{1993}).

\bibitem[{\citenamefont{Bennett et~al.}(2010)}]{Bennett:2010jb}
\bibinfo{author}{\bibfnamefont{C.~L.} \bibnamefont{Bennett}}
  \bibnamefont{et~al.} (\bibinfo{year}{2010}), \eprint{1001.4758}.

\bibitem[{\citenamefont{{Haiman} and {Knox}}(1999)}]{Haiman:1999me}
\bibinfo{author}{\bibfnamefont{Z.}~\bibnamefont{{Haiman}}} \bibnamefont{and}
  \bibinfo{author}{\bibfnamefont{L.}~\bibnamefont{{Knox}}}, in
  \emph{\bibinfo{booktitle}{Microwave Foregrounds}}, edited by
  \bibinfo{editor}{\bibnamefont{{A.~de Oliveira-Costa \& M.~Tegmark}}}
  (\bibinfo{year}{1999}), vol. \bibinfo{volume}{181} of
  \emph{\bibinfo{series}{Astronomical Society of the Pacific Conference
  Series}}, pp. \bibinfo{pages}{227--+}.

\bibitem[{\citenamefont{Dodelson}(2003)}]{Dodelson:2003ft}
\bibinfo{author}{\bibfnamefont{S.}~\bibnamefont{Dodelson}}
  (\bibinfo{year}{2003}), \bibinfo{note}{amsterdam, Netherlands: Academic
  Pr.~440 p}.

\bibitem[{\citenamefont{Challinor and Peiris}(2009)}]{Challinor:2009tp}
\bibinfo{author}{\bibfnamefont{A.}~\bibnamefont{Challinor}} \bibnamefont{and}
  \bibinfo{author}{\bibfnamefont{H.}~\bibnamefont{Peiris}},
  \bibinfo{journal}{AIP Conf. Proc.} \textbf{\bibinfo{volume}{1132}},
  \bibinfo{pages}{86} (\bibinfo{year}{2009}), \eprint{0903.5158}.

\bibitem[{\citenamefont{Kogut et~al.}(1993)}]{Kogut:1993ag}
\bibinfo{author}{\bibfnamefont{A.}~\bibnamefont{Kogut}} \bibnamefont{et~al.},
  \bibinfo{journal}{Astrophys. J.} \textbf{\bibinfo{volume}{419}},
  \bibinfo{pages}{1} (\bibinfo{year}{1993}), \eprint{astro-ph/9312056}.

\bibitem[{\citenamefont{Gordon et~al.}(2005)\citenamefont{Gordon, Hu, Huterer,
  and Crawford}}]{Gordon:2005ai}
\bibinfo{author}{\bibfnamefont{C.}~\bibnamefont{Gordon}},
  \bibinfo{author}{\bibfnamefont{W.}~\bibnamefont{Hu}},
  \bibinfo{author}{\bibfnamefont{D.}~\bibnamefont{Huterer}}, \bibnamefont{and}
  \bibinfo{author}{\bibfnamefont{T.~M.} \bibnamefont{Crawford}},
  \bibinfo{journal}{Phys. Rev.} \textbf{\bibinfo{volume}{D72}},
  \bibinfo{pages}{103002} (\bibinfo{year}{2005}), \eprint{astro-ph/0509301}.

\bibitem[{\citenamefont{Jarosik et~al.}(2010)}]{Jarosik:2010iu}
\bibinfo{author}{\bibfnamefont{N.}~\bibnamefont{Jarosik}} \bibnamefont{et~al.}
  (\bibinfo{year}{2010}), \eprint{1001.4744}.

\bibitem[{\citenamefont{{Copi} et~al.}(2006)\citenamefont{{Copi}, {Huterer},
  {Schwarz}, and {Starkman}}}]{Copietal2006}
\bibinfo{author}{\bibfnamefont{C.~J.} \bibnamefont{{Copi}}},
  \bibinfo{author}{\bibfnamefont{D.}~\bibnamefont{{Huterer}}},
  \bibinfo{author}{\bibfnamefont{D.~J.} \bibnamefont{{Schwarz}}},
  \bibnamefont{and} \bibinfo{author}{\bibfnamefont{G.~D.}
  \bibnamefont{{Starkman}}}, \bibinfo{journal}{\mnras}
  \textbf{\bibinfo{volume}{367}}, \bibinfo{pages}{79} (\bibinfo{year}{2006}),
  \eprint{arXiv:astro-ph/0508047}.

\bibitem[{\citenamefont{Hinshaw et~al.}(2009)}]{Hinshaw:2008kr}
\bibinfo{author}{\bibfnamefont{G.}~\bibnamefont{Hinshaw}} \bibnamefont{et~al.}
  (\bibinfo{collaboration}{WMAP}), \bibinfo{journal}{Astrophys. J. Suppl.}
  \textbf{\bibinfo{volume}{180}}, \bibinfo{pages}{225} (\bibinfo{year}{2009}),
  \eprint{0803.0732}.

\bibitem[{\citenamefont{{Kogut} et~al.}(1993)\citenamefont{{Kogut},
  {Lineweaver}, {Smoot}, {Bennett}, {Banday}, {Boggess}, {Cheng}, {de Amici},
  {Fixsen}, {Hinshaw} et~al.}}]{COBEDipole1993}
\bibinfo{author}{\bibfnamefont{A.}~\bibnamefont{{Kogut}}},
  \bibinfo{author}{\bibfnamefont{C.}~\bibnamefont{{Lineweaver}}},
  \bibinfo{author}{\bibfnamefont{G.~F.} \bibnamefont{{Smoot}}},
  \bibinfo{author}{\bibfnamefont{C.~L.} \bibnamefont{{Bennett}}},
  \bibinfo{author}{\bibfnamefont{A.}~\bibnamefont{{Banday}}},
  \bibinfo{author}{\bibfnamefont{N.~W.} \bibnamefont{{Boggess}}},
  \bibinfo{author}{\bibfnamefont{E.~S.} \bibnamefont{{Cheng}}},
  \bibinfo{author}{\bibfnamefont{G.}~\bibnamefont{{de Amici}}},
  \bibinfo{author}{\bibfnamefont{D.~J.} \bibnamefont{{Fixsen}}},
  \bibinfo{author}{\bibfnamefont{G.}~\bibnamefont{{Hinshaw}}},
  \bibnamefont{et~al.}, \bibinfo{journal}{\apj} \textbf{\bibinfo{volume}{419}},
  \bibinfo{pages}{1} (\bibinfo{year}{1993}), \eprint{arXiv:astro-ph/9312056}.

\bibitem[{\citenamefont{Bregman}(2009)}]{Bregman:2009qh}
\bibinfo{author}{\bibfnamefont{J.~N.} \bibnamefont{Bregman}}
  (\bibinfo{year}{2009}), \eprint{0907.3494}.

\bibitem[{\citenamefont{Law et~al.}(2009)\citenamefont{Law, Majewski, and
  Johnston}}]{Law:2009yq}
\bibinfo{author}{\bibfnamefont{D.~R.} \bibnamefont{Law}},
  \bibinfo{author}{\bibfnamefont{S.~R.} \bibnamefont{Majewski}},
  \bibnamefont{and} \bibinfo{author}{\bibfnamefont{K.~V.}
  \bibnamefont{Johnston}}, \bibinfo{journal}{Astrophys. J.}
  \textbf{\bibinfo{volume}{703}}, \bibinfo{pages}{L67} (\bibinfo{year}{2009}),
  \eprint{0908.3187}.

\bibitem[{\citenamefont{{Ghez} et~al.}(2008)\citenamefont{{Ghez}, {Salim},
  {Weinberg}, {Lu}, {Do}, {Dunn}, {Matthews}, {Morris}, {Yelda}, {Becklin}
  et~al.}}]{Ghez08}
\bibinfo{author}{\bibfnamefont{A.~M.} \bibnamefont{{Ghez}}},
  \bibinfo{author}{\bibfnamefont{S.}~\bibnamefont{{Salim}}},
  \bibinfo{author}{\bibfnamefont{N.~N.} \bibnamefont{{Weinberg}}},
  \bibinfo{author}{\bibfnamefont{J.~R.} \bibnamefont{{Lu}}},
  \bibinfo{author}{\bibfnamefont{T.}~\bibnamefont{{Do}}},
  \bibinfo{author}{\bibfnamefont{J.~K.} \bibnamefont{{Dunn}}},
  \bibinfo{author}{\bibfnamefont{K.}~\bibnamefont{{Matthews}}},
  \bibinfo{author}{\bibfnamefont{M.~R.} \bibnamefont{{Morris}}},
  \bibinfo{author}{\bibfnamefont{S.}~\bibnamefont{{Yelda}}},
  \bibinfo{author}{\bibfnamefont{E.~E.} \bibnamefont{{Becklin}}},
  \bibnamefont{et~al.}, \bibinfo{journal}{\apj} \textbf{\bibinfo{volume}{689}},
  \bibinfo{pages}{1044} (\bibinfo{year}{2008}), \eprint{0808.2870}.

\bibitem[{\citenamefont{{Gillessen} et~al.}(2009)\citenamefont{{Gillessen},
  {Eisenhauer}, {Trippe}, {Alexander}, {Genzel}, {Martins}, and
  {Ott}}}]{Gillessen09}
\bibinfo{author}{\bibfnamefont{S.}~\bibnamefont{{Gillessen}}},
  \bibinfo{author}{\bibfnamefont{F.}~\bibnamefont{{Eisenhauer}}},
  \bibinfo{author}{\bibfnamefont{S.}~\bibnamefont{{Trippe}}},
  \bibinfo{author}{\bibfnamefont{T.}~\bibnamefont{{Alexander}}},
  \bibinfo{author}{\bibfnamefont{R.}~\bibnamefont{{Genzel}}},
  \bibinfo{author}{\bibfnamefont{F.}~\bibnamefont{{Martins}}},
  \bibnamefont{and} \bibinfo{author}{\bibfnamefont{T.}~\bibnamefont{{Ott}}},
  \bibinfo{journal}{\apj} \textbf{\bibinfo{volume}{692}}, \bibinfo{pages}{1075}
  (\bibinfo{year}{2009}), \eprint{0810.4674}.

\bibitem[{\citenamefont{{Frenk} et~al.}(1988)\citenamefont{{Frenk}, {White},
  {Davis}, and {Efstathiou}}}]{1988ApJ...327..507F}
\bibinfo{author}{\bibfnamefont{C.~S.} \bibnamefont{{Frenk}}},
  \bibinfo{author}{\bibfnamefont{S.~D.~M.} \bibnamefont{{White}}},
  \bibinfo{author}{\bibfnamefont{M.}~\bibnamefont{{Davis}}}, \bibnamefont{and}
  \bibinfo{author}{\bibfnamefont{G.}~\bibnamefont{{Efstathiou}}},
  \bibinfo{journal}{\apj} \textbf{\bibinfo{volume}{327}}, \bibinfo{pages}{507}
  (\bibinfo{year}{1988}).

\bibitem[{\citenamefont{{Dubinski} and {Carlberg}}(1991)}]{1991ApJ...378..496D}
\bibinfo{author}{\bibfnamefont{J.}~\bibnamefont{{Dubinski}}} \bibnamefont{and}
  \bibinfo{author}{\bibfnamefont{R.~G.} \bibnamefont{{Carlberg}}},
  \bibinfo{journal}{\apj} \textbf{\bibinfo{volume}{378}}, \bibinfo{pages}{496}
  (\bibinfo{year}{1991}).

\bibitem[{\citenamefont{{Warren} et~al.}(1992)\citenamefont{{Warren}, {Quinn},
  {Salmon}, and {Zurek}}}]{1992ApJ...399..405W}
\bibinfo{author}{\bibfnamefont{M.~S.} \bibnamefont{{Warren}}},
  \bibinfo{author}{\bibfnamefont{P.~J.} \bibnamefont{{Quinn}}},
  \bibinfo{author}{\bibfnamefont{J.~K.} \bibnamefont{{Salmon}}},
  \bibnamefont{and} \bibinfo{author}{\bibfnamefont{W.~H.}
  \bibnamefont{{Zurek}}}, \bibinfo{journal}{\apj}
  \textbf{\bibinfo{volume}{399}}, \bibinfo{pages}{405} (\bibinfo{year}{1992}).

\bibitem[{\citenamefont{Jing and Suto}(2002)}]{Jing:2002np}
\bibinfo{author}{\bibfnamefont{Y.~P.} \bibnamefont{Jing}} \bibnamefont{and}
  \bibinfo{author}{\bibfnamefont{Y.}~\bibnamefont{Suto}},
  \bibinfo{journal}{Astrophys. J.} \textbf{\bibinfo{volume}{574}},
  \bibinfo{pages}{538} (\bibinfo{year}{2002}), \eprint{astro-ph/0202064}.

\bibitem[{\citenamefont{Kazantzidis et~al.}(2004)}]{Kazantzidis:2004vu}
\bibinfo{author}{\bibfnamefont{S.}~\bibnamefont{Kazantzidis}}
  \bibnamefont{et~al.}, \bibinfo{journal}{Astrophys. J.}
  \textbf{\bibinfo{volume}{611}}, \bibinfo{pages}{L73} (\bibinfo{year}{2004}),
  \eprint{astro-ph/0405189}.

\bibitem[{\citenamefont{Zentner et~al.}(2005)\citenamefont{Zentner, Kravtsov,
  Gnedin, and Klypin}}]{Zentner:2005wh}
\bibinfo{author}{\bibfnamefont{A.~R.} \bibnamefont{Zentner}},
  \bibinfo{author}{\bibfnamefont{A.~V.} \bibnamefont{Kravtsov}},
  \bibinfo{author}{\bibfnamefont{O.~Y.} \bibnamefont{Gnedin}},
  \bibnamefont{and} \bibinfo{author}{\bibfnamefont{A.~A.}
  \bibnamefont{Klypin}}, \bibinfo{journal}{Astrophys. J.}
  \textbf{\bibinfo{volume}{629}}, \bibinfo{pages}{219} (\bibinfo{year}{2005}),
  \eprint{astro-ph/0502496}.

\end{thebibliography}

\end{document}